\documentclass[aps,prd,reprint,twocolumn,superscriptaddress,nofootinbib,longbibliography]{revtex4-2}

\usepackage{amsmath,amssymb,bm,graphicx}
\usepackage{float}
\usepackage[letterpaper,left=0.68in,right=0.68in,top=0.62in,bottom=0.64in]{geometry}
\usepackage{hyperref}
\hypersetup{colorlinks=true,citecolor=blue,linkcolor=blue,urlcolor=blue}

\graphicspath{{figures/}}
\usepackage{changes}

\newcommand{\mpi}{m_\pi}
\newcommand{\alphaf}{\alpha_E}
\newcommand{\betaf}{\beta_M}
\newcommand{\qq}{\bm q^{\,2}}

\newcommand{\plotfile}[2][0.92\columnwidth]{%
  \IfFileExists{figures/#2}{\includegraphics[width=#1]{figures/#2}}{%
  \fbox{\parbox[c][3cm][c]{0.84\columnwidth}{\centering\footnotesize Missing: \texttt{\detokenize{#2}}}}}}

\usepackage{xcolor}

\begin{document}

\title{Reconstruction of pion polarizabilities in lattice QCD with machine learning}

\author{Hanxing Jiang}
\email{jhx@gwu.edu}
\affiliation{Physics Department, The George Washington University, Washington, DC 20052, USA}

\author{Frank X. Lee}
\email{fxlee@gwu.edu}
\affiliation{Physics Department, The George Washington University, Washington, DC 20052, USA}

\begin{abstract}
We apply the denoising diffusion probabilistic model (DDPM) from generative machine learning to a model-independent reconstruction of electric and magnetic polarizabilities of both charged and neutral pions. The inputs are the sparse momentum-space four-point-function data in lattice QCD for the connected quark-line contributions obtained in previous simulations at four pion masses between approximately 370 and 1100 MeV on  quenched Wilson lattices.
For charged pions, the elastic contribution is obtained from the pion form-factor slope, and the complete electric and magnetic inelastic terms are reconstructed separately. The neutral-pion electric and magnetic channels are reconstructed without an elastic charge-radius contribution. Four distinct families of synthetic training functions are employed, so that no single analytic fitting function is prescribed for the lattice data. 
We present low-momentum reconstructions to the limit of zero momentum transfer where the polarizabilities are defined, and mass extrapolations to the physical pion mass of 138 MeV.
\end{abstract}

\maketitle
\raggedbottom

\section{Introduction}

The electromagnetic polarizabilities of hadrons describe their response to external electric and magnetic fields~\cite{Holstein_2014}. They provide information about the internal charge and current distributions of hadrons.
The pion is of particular interest because it is the lightest meson and plays an important role in the low-energy behavior of quantum chromodynamics (QCD). Reviews of pion Compton scattering and its polarizabilities provide the broader context~\cite{Moinester:2019sew}. Its electric and magnetic polarizabilities have been studied using chiral perturbation theory (ChPT), dispersion relations, phenomenological models, and experimental measurements, see reviews in Refs.~\cite{Fil_kov_2017,Moinester:2022tba}.

Lattice QCD has a long history of  calculating magnetic moments and polarizabilities from the first principles of QCD using quark and gluon degrees of freedom. 
The standard approach is the background field method which uses weak and uniform external electromagnetic fields in a perturbative expansion~\cite{Fiebig:1988en, Burkardt_1996,Lee:2005dq,Detmold_2006,Engelhardt:2007ub,Detmold_2009,Detmold:2009fr,Alexandru:2008sj,Bignell_2019,Bali:2017ian,Bignell_2020,Luschevskaya:2015cko,Luschevskaya_2015,Tiburzi:2008ma,Davoudi:2015cba,Lujan:2014kia,Alexandru:2015dva,HuJiangTiburzi2008,Tiburzi:2008pa,Alexandru:2010dx,Lujan:2016ffj,niyazi2021charged}. For a recent review, see Ref.~\cite{Endrodi2025}.
It works best for neutral particles which remain stationary under the applied fields.
For charged particles, the acceleration under electric field and Landau levels under magnetic field complicate the extraction of polarizabilities.

To address the challenges, alternative methods are developed based on four-point correlation functions with  two electromagnetic current insertions~\cite{Wilcox_1992,BURKARDT1995441,Wilcox:1996vx,WilcoxLee2021FourPoint}. 
Even though four-point functions are more challenging to calculate on the lattice, they offer a transparent physical picture because the matrix elements  directly model Compton scattering amplitudes. Moreover, charged and neutral particles are treated on equal footing, with elastic charge-radius terms providing the difference between the two.
More lattice QCD simulations of polarizabilities are available using four-point functions~\cite{Feng:2022rkr,Lee:2023rmz,Lee2023ChargedMagnetic,Lee2024Neutral,Wang:2024,NadeemWilcoxLee2026Kaon}.

The conventional approach is to fit the lattice data with a chosen function in $\bm q^2$ and then evaluate the fit at $\bm q^2=0$. Different functions can describe the available lattice points but give different results in the region where no data are available. It was a source of significant systematic uncertainties. This is especially important when only a few momentum points are used. A good fit to the lattice data does not necessarily give a reliable zero-momentum extrapolation. It is therefore useful to explore methods that do not require choosing a single fitting function.

Machine learning has been applied to several problems in
lattice QCD~\cite{Cranmer2023MLSampling,Tomiya2025MLLQCD,kanwar2026}.
Flow-based generative models have been developed for
sampling lattice field configurations
~\cite{Albergo2019FlowBased,Kanwar2020GaugeFlows,
Boyda2021SUNFlows,Abbott2022PseudofermionFlows},
including the generation of correlated gauge-field
ensembles to reduce statistical uncertainties
~\cite{Abbott2024CorrelatedEnsembles}.
Gauge-equivariant neural networks have been used
to learn gauge-invariant observables
~\cite{Favoni2022GaugeCNN} and to improve the
numerical solution of the Dirac equation
~\cite{Cali2023DiracPreconditioners,
Lehner2023GaugePreconditioners}.
Other applications include machine-learning estimators
for lattice correlation functions
~\cite{Yoon2019MLEstimators} and the reconstruction
of spectral functions from Euclidean correlation functions
~\cite{Kades2020SpectralReconstruction,
WangShiZhou2022SpectralAD}.
Additionally, diffusion models have also been explored
for generating field configurations in lattice field
theories~\cite{WangAartsZhou2024DiffusionLFT,
Zhu2026GaugeDiffusion}.

Recently, the authors of Ref.~\cite{Alharazin_2026} applied a denoising diffusion probabilistic model (DDPM) to the reconstruction of hadronic gravitational form factors~\cite{HackettPefkouShanahan2024} which provide a related class of momentum-dependent observables based on three-point functions. The method uses synthetic functions to train a generative model, which can then reconstruct the momentum dependence from a small number of input points. Instead of fitting the lattice data with one prescribed function, the DDPM generates functions based on the momentum dependencies learned during training. This provides an alternative approach to the extrapolation problem. For a pedagogical introduction to diffusion models in machine learning, see Ref.~\cite{lai2026}.

In this work, we apply the DDPM method to the reconstruction of pion electromagnetic polarizabilities from four-point-function lattice data generated in previous studies~\cite{Lee:2023rmz,Lee2023ChargedMagnetic,Lee2024Neutral}. We use four families of training functions: multipole, $z$-expansion, meson-dominance, Pad\'e, and modified exponential functions. These functions provide different momentum dependences for training the model. The reconstruction does not assume that the lattice data follow any one of these functions. 
We first reconstruct the charged-pion electromagnetic form factor at four pion masses.
The derivative of the form factor at zero momentum gives the charge radius and the corresponding elastic contributions to the electric and magnetic polarizabilities. We then reconstruct the electric and magnetic inelastic contributions from the lattice inputs. The elastic and inelastic results at $\qq=0$ are combined to obtain the total polarizabilities at each pion mass. The same method is applied to the neutral pion, where elastic contributions are absent. 
Then a subsequent chiral extrapolation is performed to bring the polarizabilities to the physical pion mass $m_\pi=138$ MeV, and compared with the results from conventional fitting methods and available theoretical and experimental results. 

The paper is organized as follows. In Sec.~II, we review the four-point-function method and describe the training functions and DDPM reconstruction. Section~III presents the lattice data and numerical results. 
Section~IV contains the conclusion and outlook. Additional details of the numerical implementation, derivative extraction, and reconstruction diagnostics are given in the appendices.

\section{Methodology}

\subsection{Four-point functions and polarizabilities}

The four-point-function method relates the low-energy Compton amplitude to Euclidean matrix elements of two electromagnetic currents between pion sources, following the formulation in Ref.~\cite{WilcoxLee2021FourPoint}. In the zero-momentum Breit frame, the pion source and sink are at rest and the currents carry opposite spatial momenta $\bm q$. We use $\qq=|\bm q|^2$ throughout.

The use of spatial charge correlations predates the present four-point polarizability formulation. Earlier lattice work explored the elastic charge-overlap limit~\cite{Wilcox_1992,Andersen:1996qb} and its implications for charged-pion polarizability~\cite{Wilcox:1996vx}. The interpretation of hadron correlation functions has been studied in lattice QCD~\cite{BURKARDT1995441}, as have continuum-moment relations on the lattice~\cite{Wilcox_2002}. Two-current correlations in the pion have also been calculated directly~\cite{Bali_2018}. These are related studies, rather than additional data inputs to the present reconstruction.

For a charged pion, the electric and magnetic polarizabilities are separated into elastic and inelastic contributions. With $\alpha$ the fine-structure constant and $r_E$ the charge radius, the expressions used in the original four-point analysis~\cite{Lee:2023rmz,Lee2023ChargedMagnetic} are
\begin{align}
\alphaf^{\pi^\pm}
&=\frac{\alpha \langle  r_E^2\rangle}{3\mpi}
+\lim_{\qq\to0}\frac{2\alpha}{\qq}\int_0^\infty dt\,
\left[Q_{44}(\bm q,t)-Q_{44}^{\rm elas}(\bm q,t)\right],
\label{eq:charged_e}\\
\betaf^{\pi^\pm}
&=-\frac{\alpha \langle  r_E^2\rangle}{3\mpi}
+\lim_{\qq\to0}\frac{2\alpha}{\qq}\int_0^\infty dt\,
\left[Q_{11}^{\rm inel}(\bm q,t)-Q_{11}^{\rm inel}(\bm 0,t)\right].
\label{eq:charged_m}
\end{align}
Note that the charge radius term appears in both formulas with opposite signs.
For the neutral pion, the charge-radius elastic term is absent. The corresponding connected-contribution expressions~\cite{Lee2024Neutral} are
\begin{align}
\alphaf^{\pi^0}
&=\lim_{\qq\to0}\frac{2\alpha}{\qq}\int_0^\infty dt\,
Q_{44}(\bm q,t),
\label{eq:neutral_e}\\
\betaf^{\pi^0}
&=\lim_{\qq\to0}\frac{2\alpha}{\qq}\int_0^\infty dt\,
\left[Q_{11}(\bm q,t)-Q_{11}(\bm 0,t)\right].
\label{eq:neutral_m}
\end{align}

The four-point function $Q_{\mu\mu}$ is defined as,
\fontsize{9.5}{9.5}
\begin{align}
&Q_{\mu\mu}(\bm q,t_3,t_2,t_1,t_0)  \equiv  \label{eq:Qmm} \\&
 \frac{ \displaystyle\sum_{\bm x_3,\bm x_2,\bm x_1,\bm x_0} e^{-i\bm q\cdot \bm x_2} e^{i\bm q\cdot \bm x_1} 
\langle \Omega | \psi (x_3) :j^L_\mu(x_2) j^L_\mu(x_1):  \psi^\dagger (x_0) |\Omega \rangle }
{\displaystyle\sum_{\bm x_3,\bm x_0} \langle \Omega  | \psi (x_3) \psi^\dagger (x_0) |\Omega \rangle }. \nonumber
\end{align}
In this expression, $\Omega$ denotes the vacuum, 
$\psi$ the interpolating field of the pion and $j^{L}_\mu$ the lattice version of 
the electromagnetic current density. The two-point function in the denominator is for normalization. 
Normal ordering is used to include the required subtraction of vacuum expectation values (VEV) on the lattice. 
The sums over $\bm x_0$ and $\bm x_3$ enforce zero-momentum pions at the source ($t_0$) and sink ($t_3$). 
The sum over $\bm x_1$ injects momentum $\bm q$ by the current at $t_1$, 
whereas sum over $\bm x_2$ takes out $\bm q$ by the current at $t_2$ to satisfy energy-momentum conservation in the process.
The two possibilities of time ordering are implied in the normal ordering. The time $t$ in the formulas represents the separation between the two currents $t=t_2-t_1$ with the two fixed ends $t_0$ and $t_3$ implied.

The elastic piece in the four-point function has information on the form factor of the pion through the amplitude squared at large time separations,
\begin{equation}
Q^{elas}_{44}(\bm q,t)= {(E_\pi+m_\pi)^2\over 4 E_\pi m_\pi} F^2_\pi(Q^2) \, e^{-a(E_\pi-m_\pi) t}.
\label{eq:Q44elas}
\end{equation}
The charge radius $ \langle  r_E^2\rangle $ in the formula can then be extracted from  $F_\pi$. Note that the pion form factor is a function of four-momentum transfer $Q^2$ (in space-like region $Q^2>0$) defined by the four-momentum transfer squared in Minkowski spacetime,
\begin{equation}
Q^2=-(E_\pi(\bm q)-E_\pi(\bm 0))^2+ \bm q^2=2m_\pi(E_\pi-m_\pi),
\label{eq:Q2}
\end{equation}
where $E_\pi(\bm q) = \sqrt{\bm q^2 +m_\pi^2}$.
Only the form factor $F_\pi(Q^2)$ in Eq.\eqref{eq:Q44elas} is treated as a function of $Q^2$; the rest of the four-point function formalism is a function of three-momentum transfer $\bm q$. The $Q^2$ is also a function of  $\bm q^2$ and both reach the limit of zero simultaneously.

\begin{figure}
\centering
\includegraphics[width=0.36\textwidth]{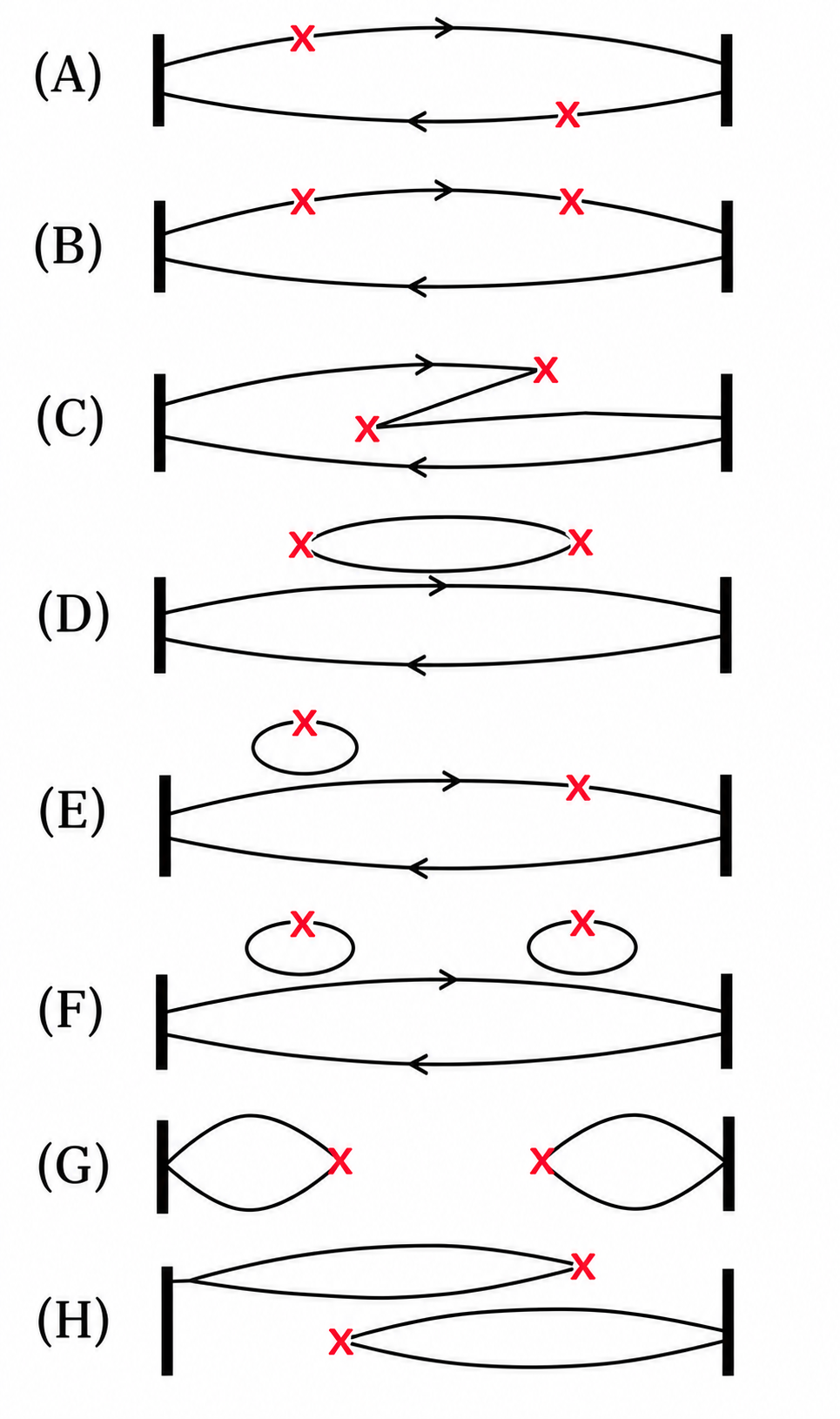}
\caption{Quark-line topologies contributing to the pion four-point function. Vertical bars denote pion source and sink, arrows indicate quark propagation, and crosses denote electromagnetic-current insertions. Diagrams (A)--(F) are connected and disconnected contributions to both charged and neutral pions. 
Additional disconnected diagrams (G)--(H) contribute to neutral pions.
}
\label{fig:diagrams}
\end{figure}

At the quark level, Wick contractions of quark-
antiquark pairs in Eq.\eqref{eq:Qmm} lead to distinct quark-line diagrams illustrated in Fig.~\ref{fig:diagrams}.
Diagrams (A)--(C) are connected quark-line contractions, while (D)--(H) contain disconnected quark loops. The original simulations and the present reconstruction include only (A)--(C). Disconnected contributions are not included in the lattice inputs analyzed here. 

The quantities supplied for DDPM reconstruction in the charged electric and magnetic inelastic channels are the \emph{entire second terms} of Eqs.~(\ref{eq:charged_e}) and (\ref{eq:charged_m}), already including their original prefactors and momentum dependence. No additional division by $\qq$ or decomposition of these processed inputs is performed.

The normalized and processed lattice inputs for the neutral channels are used directly. Since disconnected contractions have not been evaluated, the resulting quantities are connected-contribution polarizabilities rather than complete QCD predictions.

\subsection{Training functions}
\label{sec:fun}

The DDPM training data are generated from four families of smooth functions: multipole, $z$-expansion, meson-dominance, and Pad\'e approximants. These functions provide different momentum dependences for training the model. They are not used as separate fitting functions for the lattice data. We use $x$ as the momentum-squared variable, with $x=Q^2$ for the elastic form factor and $x=\bm q^2$ for the inelastic contributions. The condition $F_\pi(0)=1$ is imposed separately during the reconstruction of the charged-pion form factor, while no such condition is imposed on the inelastic contributions.

\subsubsection{Multipole functions}

The multipole family describes the momentum dependence through a pole-like function,
\begin{equation}
f_{\rm mp}(x)
=A\left(1+\frac{x}{\Lambda^2}\right)^{-n}.
\label{eq:multipole}
\end{equation}
The parameter $A$ determines the normalization, $\Lambda$ sets the momentum scale, and $n$ controls the falloff and curvature. The derivative at zero momentum is $f'_{\rm mp}(0)=-An/\Lambda^2$. This family provides a simple description of form-factor-like behavior, although the slope and curvature are related through the same parameters.

\subsubsection{The $z$ expansion}

The $z$ expansion uses a conformal mapping to describe the analytic momentum dependence of form factors. It has been applied to charge-radius studies~\cite{HillPaz2010} and heavy-meson transition form factors~\cite{BourrelyCapriniLellouch2009}. We use
\begin{equation}
z(x)=
\frac{\sqrt{t_{\rm cut}+x}-\sqrt{t_{\rm cut}+x_0}}
{\sqrt{t_{\rm cut}+x}+\sqrt{t_{\rm cut}+x_0}},
\label{eq:zvariable}
\end{equation}
where $x_0=-t_0$ is the reference momentum squared, and $t_{\rm cut}$ specifies the branch point. For the pion electromagnetic form factor, the physical two-pion threshold is $t_{\rm cut}=4m_\pi^2$. In generating the synthetic training curves, however, $t_{\rm cut}$ and $x_0$ are varied to provide different momentum dependences. The functions are expressed as
\begin{equation}
f_z(x)=\sum_{k=0}^{K}a_k[z(x)]^k.
\label{eq:zfamily}
\end{equation}
The coefficients $a_k$ determine the shape of the curves, allowing different slopes and curvatures near zero momentum. Unlike a single pole function, the expansion can describe a wider range of momentum dependences through its higher-order terms.

\subsubsection{Meson-dominance functions}

The meson-dominance family uses a sum of pole terms to describe the momentum dependence. Such forms are related to meson-pole descriptions of hadronic form factors, including the vector-meson-dominance approach~\cite{Sakurai1969,GounarisSakurai1968}. We use
\begin{equation}
f_{\rm MD}(x)=
\sum_{j=1}^{N_p}\frac{c_j}{m_j^2+x}.
\label{eq:mesondom}
\end{equation}
The masses $m_j$ determine the momentum scales, while the coefficients $c_j$ give the relative contributions of the poles. In the training functions, the pole masses are sampled from selected meson masses and additional effective mass values, without restricting them to vector mesons. The sum allows several momentum scales to contribute to the slope and curvature, providing more flexibility than a single multipole function.

\subsubsection{Pad\'e approximants}

The fourth family consists of rational functions known as Pad\'e approximants. Such functions have been used in extrapolation and approximation methods~\cite{Schlessinger1968,MasjuanPeris2010}. We use
\begin{equation}
f_{\rm P}(x)=
\frac{a_0+\sum_{i=1}^{N_a}a_i x^i}
{1+\sum_{j=1}^{N_b}b_j x^j}.
\label{eq:padefamily}
\end{equation}
Here $N_a$ and $N_b$ are the highest powers in the numerator and denominator, respectively. The coefficients are sampled to generate different momentum dependences, with the denominator restricted to avoid poles in the training interval. The rational form allows the slope and curvature to vary without requiring a single pole mass. Together with the other three families, these functions provide a range of shapes for the DDPM reconstruction.

\subsection{Denoising diffusion probabilistic model (DDPM)}

Diffusion-based generative modeling was introduced using
nonequilibrium thermodynamics~\cite{SohlDickstein2015},
followed by the denoising diffusion probabilistic model
(DDPM)~\cite{Ho2020DDPM}. The training data are generated
from the four function families described in the previous
section. Each training curve is represented by a vector
$\bm x_0\in\mathbb R^{200}$ on a uniform grid of 200 points
over $0\leq x\leq2\,{\rm GeV}^2$, where $x$ denotes $Q^2$
for the elastic form factor and $\bm q^2$ for the inelastic
contributions. The grid covers the range of $Q^2$ in the
form factor $F_\pi(Q^2)$ for all pion masses considered
in this work.

The DDPM consists of a forward diffusion process and a
reverse denoising process. In the forward process, Gaussian
noise is gradually added to a training curve. At step $s$,
the noisy curve is given by
\begin{equation}
\bm x_s=
\sqrt{\bar\alpha_s}\bm x_0+
\sqrt{1-\bar\alpha_s}\bm\epsilon,
\quad
\bm\epsilon\sim\mathcal N(0,\bm I),
\label{eq:forward}
\end{equation}
where $\bar\alpha_s=\prod_{j=1}^s(1-\beta_j)$ and $\beta_j$
controls the amount of noise added at each step. We use
a cosine noise schedule~\cite{NicholDhariwal2021}
and the $v$-prediction method~\cite{SalimansHo2022Distillation},
with the training target defined as
$\bm v_s=\sqrt{\bar\alpha_s}\bm\epsilon
-\sqrt{1-\bar\alpha_s}\bm x_0$.
The neural network is trained by minimizing
\begin{equation}
\mathcal L_{\rm DDPM}
=
\mathbb E_{\bm x_0,s,\bm\epsilon}
\left[
\left\|
\bm v_s-
\bm v_\theta(\bm x_s,s)
\right\|^2
\right],
\label{eq:ddpm_loss}
\end{equation}
where $\bm v_\theta$ is the network prediction.
The predicted noise can be recovered from
$\bm\epsilon_\theta=
\sqrt{1-\bar\alpha_s}\bm x_s+
\sqrt{\bar\alpha_s}\bm v_\theta$.
Once trained, the model generates new curves by reversing
the diffusion process.

For reconstruction, the lattice data are used to constrain
the generated curves at the available momentum points.
This is related to diffusion-based methods for inverse
problems~\cite{Daras2024DiffusionSurvey}. The trained model
supplies the momentum dependence learned from the training
functions, while the lattice data provide the constraints.
The curves can then be reconstructed in the low-momentum
region where no lattice data are available. The derivative
of the pion form factor is determined at $Q^2=0$, while
the inelastic contributions are evaluated at $\qq=0$.
Details of the training procedure and numerical
implementation are given in Appendix~\ref{app:DDPM}.

For a given set of lattice data, we generate $N$
reconstructed curves. We take the pointwise median of
these curves as the central reconstruction,
\begin{equation}
f_{\rm med}(x_i)=
\mathop{\rm median}_{k=1,\ldots,N}f_k(x_i).
\label{eq:median}
\end{equation}
The median curve is used to extract the derivative at
$Q^2=0$ for the charge radius, or the value at $\qq=0$
for the inelastic contributions.

We do not use the spread of the $N$ generated curves as
the statistical error, since this spread depends on the
choice of training functions. Adding another function
family can increase the spread even when the lattice
input and the median reconstruction remain nearly
unchanged. The spread therefore does not directly
represent the statistical uncertainty of the lattice data.

Instead, we generate $M$ sets of lattice inputs by Gaussian
sampling of the measured points, using their reported
jackknife errors as the standard deviations. For each set,
we generate $N$ curves and calculate the physical quantity
from their pointwise median. The sample standard deviation
of the $M$ extracted values is used as the statistical
error, while the central value is obtained from the
original lattice measurements. The quoted errors do not
include possible systematic uncertainties from the choice
of training functions. We discuss this limitation in
Sec.~\ref{sec:sum}.

\section{Simulation Details and Results}

\subsection{Lattice data and numerical setup}

The lattice inputs originate from the four-point-function studies of charged-pion electric~\cite{Lee:2023rmz}, charged-pion magnetic~\cite{Lee2023ChargedMagnetic}, and neutral-pion polarizabilities~\cite{Lee2024Neutral}. Those calculations used a quenched Wilson gauge ensemble of 1000 configurations on a $24^3\times48$ lattice at gauge coupling $\beta=6.0$ and four hopping parameters $\kappa=0.1520$, $0.1543$, $0.1555$, and $0.1565$, corresponding to pion masses of approximately 1100, 800, 600, and 370 MeV. They used conserved electromagnetic currents and evaluated the connected Wick contractions. We analyze the processed momentum-space lattice results. 

Four nonzero spatial momenta $\bm q=(0,0,1)$, $(0,1,1)$, $(1,1,1)$, and $(0,0,2)$ were considered in the lattice simulations. The corresponding $\qq$ and $Q^2$ from Eq.\eqref{eq:Q2} in physical units are given in Table~\ref{tab:lattice_input} along with all the input lattice data. 
The charged magnetic inelastic input doe not have the $\bm q=(1,1,1)$ point to avoid elastic contribution in the second term in Eq.\eqref{eq:charged_m} because the spatial current component $j_1$ is inserted in the four-point functions~\cite{Lee2023ChargedMagnetic}.  The exact lattice central values and statistical uncertainties are taken from the corresponding processed datasets.

The baseline DDPM uses a grid of 200 points and the four training-function families described in Section~\ref{sec:fun}. The standard reconstruction uses $M=100$ lattice resamplings to propagate lattice errors and $N=1000$ generated curves per realization for the median. The generated-ensemble width is not used for the reported statistical uncertainty. All shaded bands in the reconstruction figures represent propagated lattice-data errors. The numerical extraction below uses the median curves.
\begin{table*}
\centering
\caption{Lattice input data used for DDPM reconstruction at the four pion masses. They are taken from~\cite{Lee:2023rmz,Lee2023ChargedMagnetic,Lee2024Neutral}. The quoted errors are statistical only from a jackknife analysis. The four spatial momenta $\bm q=(0,0,1)$, $(0,1,1)$, $(1,1,1)$, and $(0,0,2)$ correspond to $\bm q^2=0.366$, $0.733$, $1.100$, and $1.470$ GeV$^2$, respectively. 
The corresponding $Q^2$ for each pion mass is obtained via Eq.\eqref{eq:Q2}. The form factor is given as a function of $Q^2$, while the polarizability inputs are functions of $\bm q^2$. The $\bm q=(1,1,1)$ point is excluded in the charged magnetic inelastic reconstruction (see reason in main text).}
\label{tab:lattice_input}

\footnotesize
\setlength{\tabcolsep}{2pt}
\renewcommand{\arraystretch}{1.15}


\begin{tabular}{c|cccc|cccc}
\hline
&
\multicolumn{4}{c|}{$Q^2$ (GeV$^2$)}
& \multicolumn{4}{c}{$F_\pi(Q^2)$} \\
\cline{2-9}
$m_\pi$ (MeV)
& $(0,0,1)$ & $(0,1,1)$ & $(1,1,1)$ & $(0,0,2)$
& $(0,0,1)$ & $(0,1,1)$ & $(1,1,1)$ & $(0,0,2)$ \\
\hline
1100
& 0.342 & 0.646 & 0.923 & 1.180
& 0.8209(23) & 0.7213(23) & 0.6500(37) & 0.6043(46) \\
800
& 0.325 & 0.595 & 0.830 & 1.040
& 0.7677(27) & 0.6462(59) & 0.5681(101) & 0.5522(115) \\
600
& 0.303 & 0.534 & 0.730 & 0.901
& 0.7412(15) & 0.6360(25) & 0.5834(36) & 0.5255(117) \\
370
& 0.251 & 0.416 & 0.549 & 0.663
& 0.7200(36) & 0.6161(45) & 0.5535(62) & 0.5297(83) \\
\hline
\end{tabular}

\par\vspace{4pt}


\begin{tabular}{c|cccc|cccc}
\hline
&
\multicolumn{4}{c|}{$-\alpha_E^{\pi^\pm,\mathrm{inel}}(\bm q^2)$}
& \multicolumn{4}{c}{$\beta_M^{\pi^\pm,\mathrm{inel}}(\bm q^2)$} \\
\cline{2-9}
$m_\pi$ (MeV)
& $(0,0,1)$ & $(0,1,1)$ & $(1,1,1)$ & $(0,0,2)$
& $(0,0,1)$ & $(0,1,1)$ & $(1,1,1)$ & $(0,0,2)$ \\
\hline
1100
& 0.1907(43) & 0.1002(15) & 0.0445(10) & 0.0224(7)
& 0.5443(12) & 0.4049(6) & --- & 0.2519(2) \\
800
& 0.3455(52) & 0.1243(25) & 0.0365(18) & 0.0382(14)
& 0.9192(20) & 0.6424(9) & --- & 0.3786(4) \\
600
& 0.5765(31) & 0.2911(9) & 0.1728(4) & 0.1140(2)
& 1.3734(32) & 0.9209(14) & --- & 0.5265(7) \\
370
& 0.7176(82) & 0.2966(33) & 0.1395(21) & 0.0852(16)
& 2.1914(72) & 1.4319(34) & --- & 0.8020(16) \\
\hline
\end{tabular}

\par\vspace{4pt}


\begin{tabular}{c|cccc|cccc}
\hline
&
\multicolumn{4}{c|}{$\alpha_E^{\pi^0}(\bm q^2)$}
& \multicolumn{4}{c}{$\beta_M^{\pi^0}(\bm q^2)$} \\
\cline{2-9}
$m_\pi$ (MeV)
& $(0,0,1)$ & $(0,1,1)$ & $(1,1,1)$ & $(0,0,2)$
& $(0,0,1)$ & $(0,1,1)$ & $(1,1,1)$ & $(0,0,2)$ \\
\hline
1100
& 0.5278(23) & 0.3788(7) & 0.2777(3) & 0.2201(2)
& 0.6536(14) & 0.2588(7) & 0.0548(4) & 0.0350(3) \\
800
& 0.4215(21) & 0.2822(6) & 0.2125(2) & 0.1740(1)
& 0.5766(21) & 0.1855(10) & 0.0032(6) & $-0.0093(5)$ \\
600
& 0.3183(20) & 0.2116(6) & 0.1714(2) & 0.1490(1)
& 0.4508(31) & 0.1135(14) & $-0.0454(9)$ & $-0.0493(7)$ \\
370
& 0.0841(26) & 0.1377(8) & 0.1501(4) & 0.1460(2)
& 0.3025(77) & $-0.0035(36)$ & $-0.1610(23)$ & $-0.1419(17)$ \\
\hline
\end{tabular}

\end{table*}

\subsection{Charged-pion elastic contribution}
\begin{figure}
\centering
\includegraphics[width=0.45\textwidth]{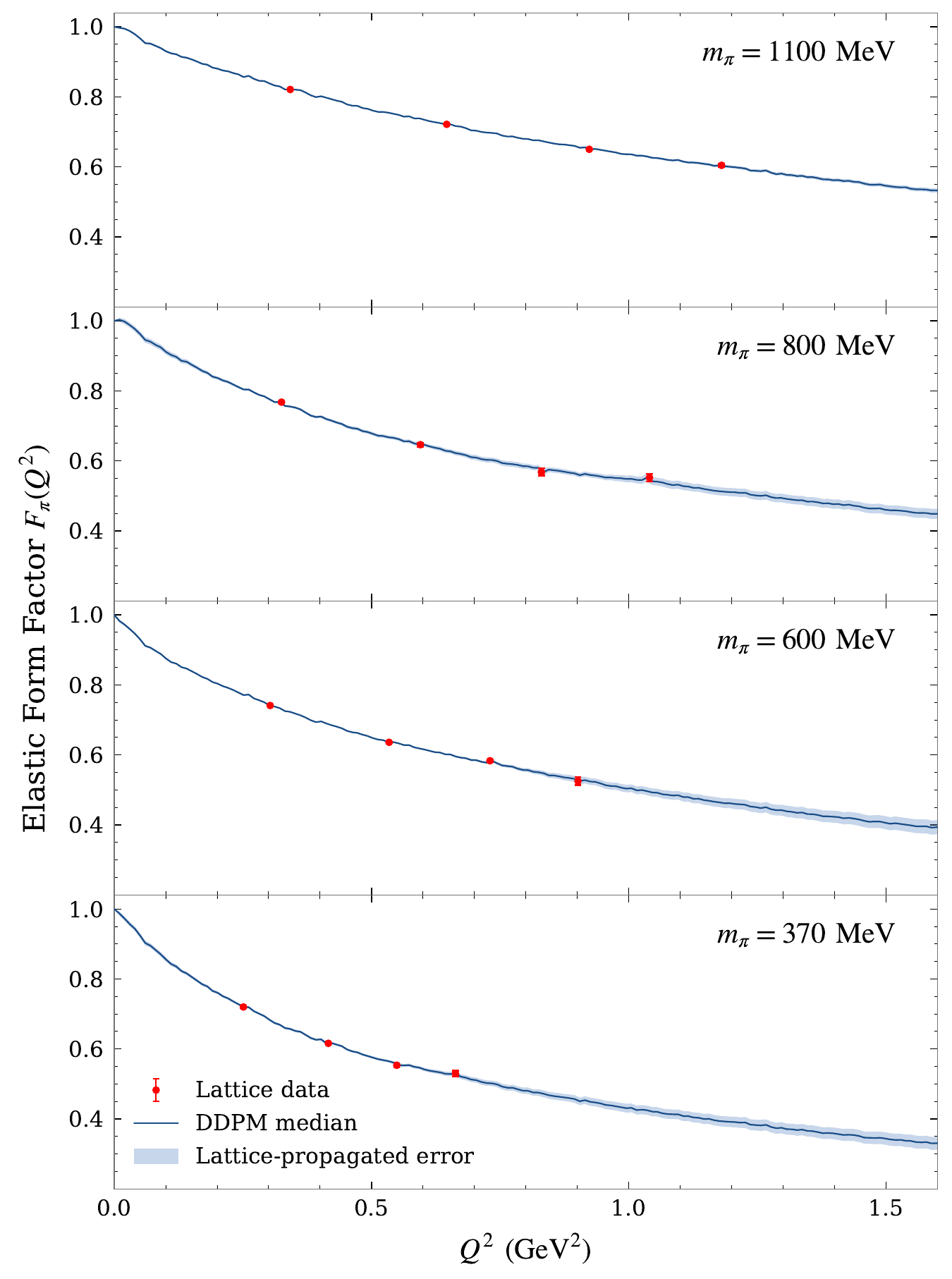}
\caption{Reconstructed charged-pion electromagnetic form factor $F_\pi(Q^2)$ at the four pion masses.
The red points are lattice input, the solid curves are the DDPM-reconstructed medians, and the shaded bands represent the statistical uncertainties propagated from the lattice data. All curves are reconstructed with the constraint $F_\pi(0)=1$.}
\label{fig:elastic_ff}
\end{figure}

We first apply the DDPM to the pion electromagnetic form factor. Figure~\ref{fig:elastic_ff} shows the reconstructed form factors at the four pion masses. 
The four-point function formalism for pion electromagnetic polarizabilities has access to the pion  form factor $F_\pi(Q^2)$ via Eq.\eqref{eq:Q44elas}. 
The charge radius can be accessed via its low-momentum expansion,
\begin{equation}
F_\pi(Q^2)=1-\frac{\langle r_E^2\rangle}{6}Q^2
+\cdots,
\quad
\langle r_E^2\rangle=-6\left.\frac{dF_\pi}{dQ^2}\right|_{Q^2=0}.
\label{eq:radius}
\end{equation}

The elastic contribution in Eq.\eqref{eq:charged_e} is determined by the charge radius, which requires the derivative of the form factor at zero momentum. We extract the derivative by fitting the reconstructed form factor with a quadratic function in the low-momentum region. The coefficient of the linear term gives the derivative at zero momentum transfer. However, the result depends on the fitting interval used. A small interval can give large fluctuations in the derivative, while a larger interval includes more points away from zero momentum and may affect the extracted slope.

To determine a suitable fitting interval, we examine the derivative as the upper end of the interval is varied. We find that the derivative first fluctuates, then reaches a relatively stable region, and finally changes as more higher-momentum points are included. We select the intermediate stable region for the derivative extraction. The fitting intervals are determined separately for the four pion masses. The detailed analysis and the selected windows are given in Appendix~\ref{app:deriv}.

\subsection{Charged-pion electric inelastic contribution}
\label{sec:800mev}

\begin{figure}
\centering
\includegraphics[width=0.45\textwidth]{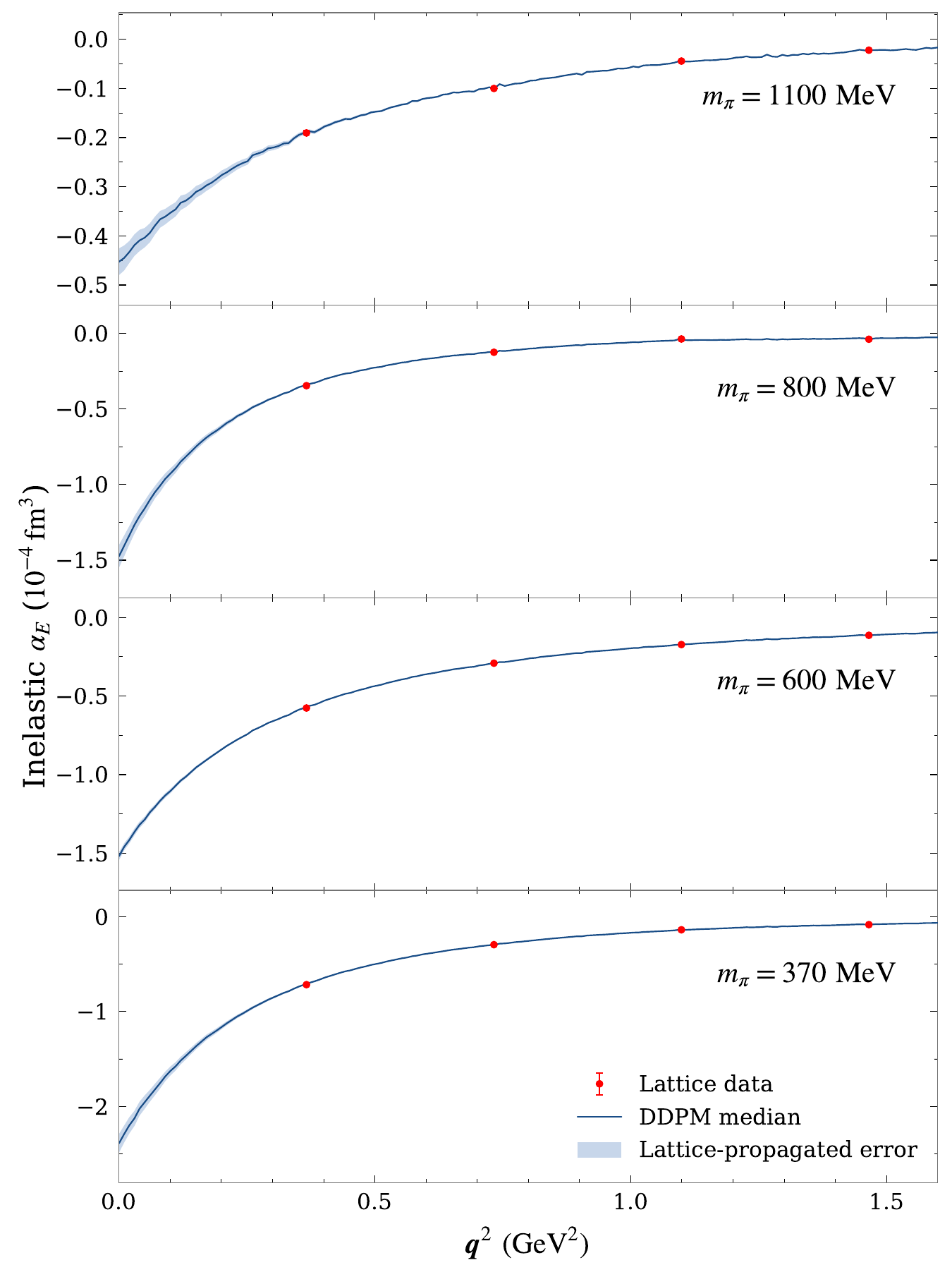}
\caption{Reconstruction of charged-pion electric inelastic contribution as a function of $\qq$ at the four pion masses. Red symbols are processed lattice inputs, blue curves are DDPM medians, and pale bands show lattice-propagated statistical uncertainty. The intercept at $\qq=0$ determines $\alpha_E^{\rm inel}$, the 2nd term in Eq.\eqref{eq:charged_e}.}
\label{fig:charged_einel}
\end{figure}

We next consider the electric inelastic contribution. The lattice input for the DDPM is the complete second term of Eq.\eqref{eq:charged_e}. It includes the momentum factors and other coefficients in the original polarizability formula. We reconstruct this term directly, without separating or modifying its factors. Its value at $\qq=0$ gives the electric inelastic polarizability (denoted as $\alpha_E^{\rm inel}$).
Figure~\ref{fig:charged_einel} shows the reconstructed electric inelastic contributions at the four pion masses.  
We see that the electric inelastic contribution is a smooth decreasing function of $\bm q^2$, negative at all four pion masses. 
The reconstruction extends the momentum dependence to $\qq=0$, where no direct lattice data are available. It is this extrapolated value at $\qq=0$ that yields the inelastic contribution to the polarizability.  This is different from the form factor reconstruction where it is constrained by $F_\pi(0)=1$. 

The DDPM reconstruction in this channel presents an unexpected challenge. 
The reconstructed curve at $m_\pi=800$ MeV in Figure~\ref{fig:charged_einel} exhibits an unusual behavior: it takes a deeper dive than at the other masses in the small $\bm q^2$ region. 
This was not observed in the original work (see Fig.12 of Ref.~\cite{Lee:2023rmz}).
To examine this behavior, we repeat the DDPM reconstruction using different subsets of the four lattice points. The results show that the low-momentum extrapolation is sensitive to the choice of input points. 
We found that the reconstruction using the second and fourth points gives a result more consistent with the mass dependence observed at the other pion masses.
The comparison with the original four-point reconstruction and the other two- and three-point combinations is detailed in Appendix~\ref{app:800diagnostic}.

This example shows an advantage of the DDPM method. The reconstruction can be performed with only a few lattice momentum points, and different combinations of the input data can be used to examine their consistency. In this case, the comparison helps identify the unusual behavior of the 800-MeV data, especially the third point. The agreement with the other pion masses does not by itself establish the correct physical result, but it provides a useful check on the lattice input and indicates which points require further examination.

\subsection{Charged-pion magnetic inelastic contribution}

The magnetic inelastic contribution is reconstructed in the same way as the electric inelastic contribution. The DDPM input is the complete second term of Eq.\eqref{eq:charged_m}, including the factors already contained in the lattice result. The value of the reconstructed curve at $\qq=0$ gives $\beta_M^{\rm inel}$.

\begin{figure}
\centering
\includegraphics[width=0.45\textwidth]{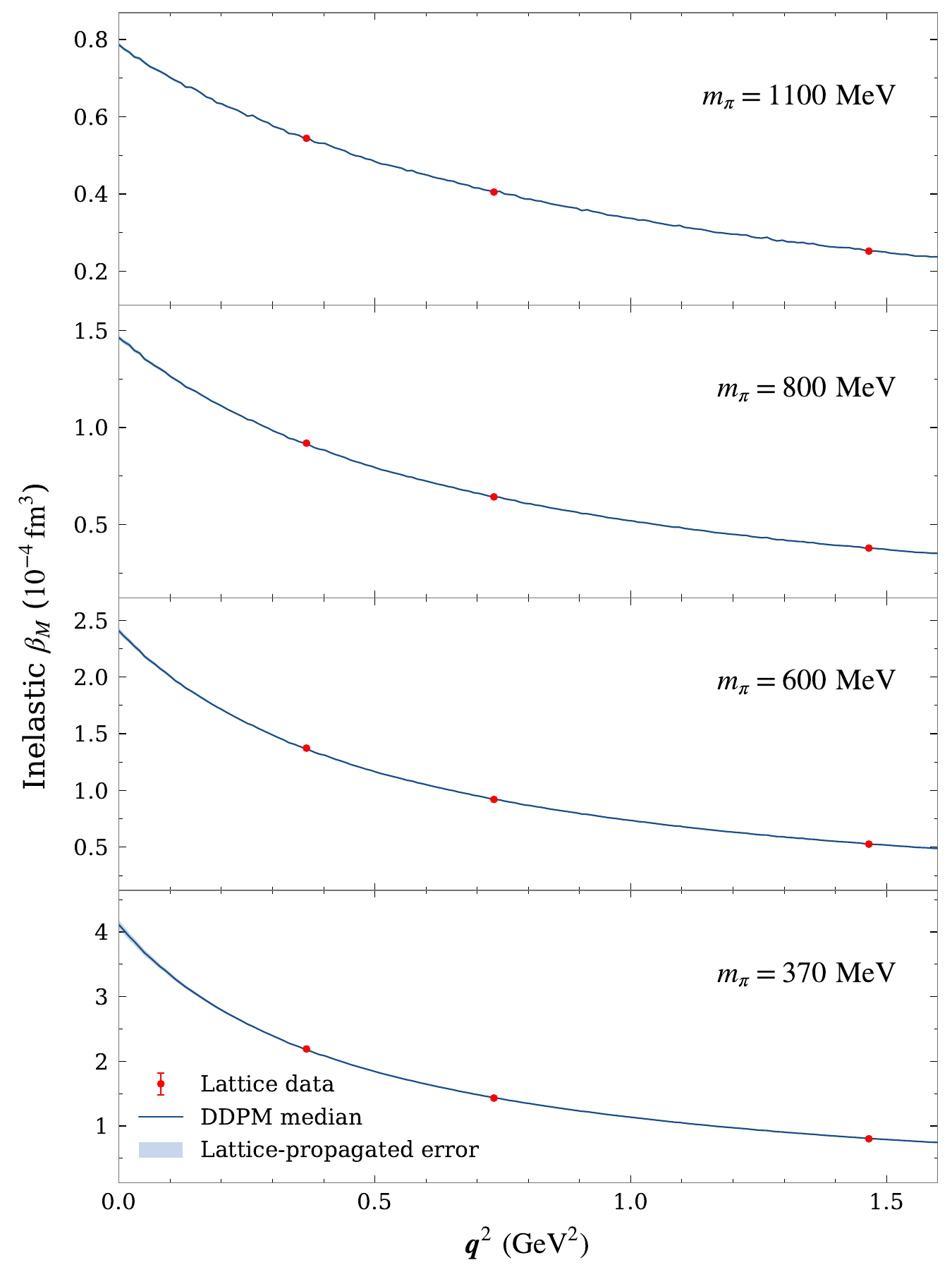}
\caption{Reconstruction of charged-pion magnetic inelastic contribution as a function of $\qq$ at the four pion masses. Only three nonzero momentum points are available from the lattice input. The $\qq\to0$ intercept gives $\beta_M^{\rm inel}$, the 2nd term in Eq.\eqref{eq:charged_m}.}
\label{fig:charged_minel}
\end{figure}

For the magnetic reconstruction, we only have three nonzero momentum points $\bm q=(0,0,1)$, $(0,1,1)$, $(0,0,2)$ from the lattice input. The $\bm q=(1,1,1)$ point is not available 
because the x-component of the current $j^L_1$ is used to construct the magnetic four-point functions~\cite{Lee2023ChargedMagnetic}. This is not a problem for the DDPM reconstruction, as shown in Figure~\ref{fig:charged_minel}.  
We see that the magnetic inelastic contribution is a smooth rising function of $\bm q^2$, positive at all four pion masses. The extrapolated values at $\bm q^2=0$ from the DDPM reconstruction form a smooth function of pion mass with small error bars.
This is another example of the DDPM feature: a reliable reconstruction can be achieved even with incomplete input data. The full dependence can be recovered with as few as two well-placed lattice data points.

\subsection{Neutral-pion polarizabilities}

\begin{figure}
\centering
\includegraphics[width=0.45\textwidth]{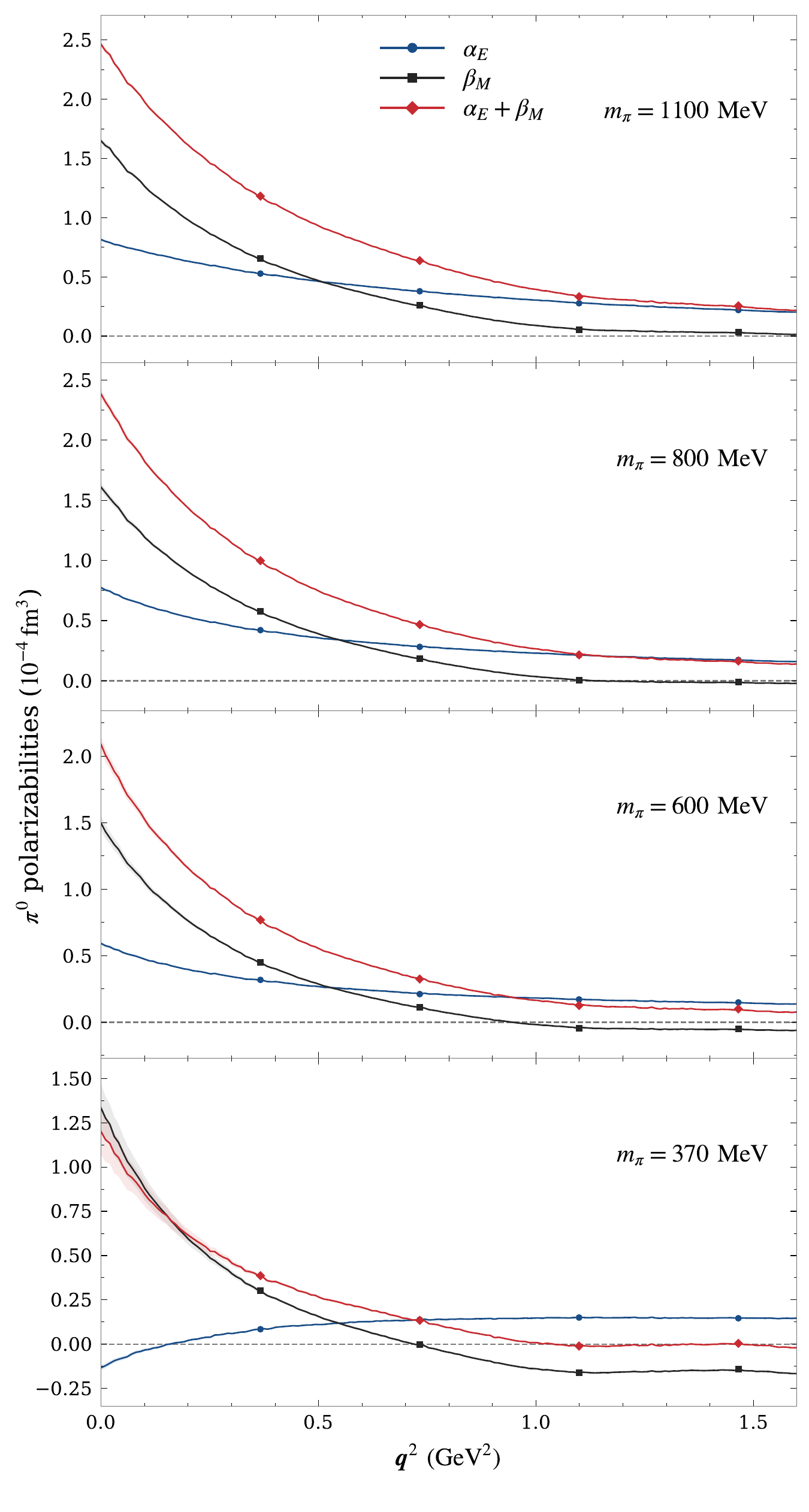}
\caption{Reconstruction of neutral-pion polarizabilities at the four pion masses. 
The three curves correspond to $\alpha_E$ (blue), $\beta_M$ (black), and $\alpha_E+\beta_M$ (red).
}
\label{fig:neutral_recon}
\end{figure}

Here we consider the electric and magnetic polarizabilities of the neutral pion for which the lattice data are given in Table~\ref{tab:lattice_input}. Unlike the charged pion, there is no elastic charge-radius contribution, see Eq.\eqref{eq:neutral_e} and Eq.\eqref{eq:neutral_m}. 
Figure~\ref{fig:neutral_recon} shows the reconstructed neutral-pion electric and magnetic polarizabilities at the four masses. The sum of the two polarizabilities is also shown, as is done in the original work~\cite{Lee2024Neutral}. 
The polarizabilities are obtained from the values at $\bm q^2=0$.

An interesting feature in the $\bm q^2$ dependence of neutral pion polarizabilities occurs when the lattice input is near zero or crosses zero in the momentum region considered, as found in the $\beta_M$ at 370 and 600 MeV. The training functions do not cover this behavior equally well. In particular, the multipole functions in Eq.\eqref{eq:mesondom} do not cross zero. A direct DDPM reconstruction of such data can therefore give an unstable result.
We found a simple remedy for this problem. We shift the lattice input upward by a constant before reconstruction. The DDPM is then applied to the shifted data. After obtaining the reconstructed curve, we subtract the same constant to recover the original quantity. The shift does not change the lattice errors and is removed before the polarizability is extracted. The details of this analysis can be found in Appendix~\ref{app:neutral_shift}.

We emphasize that the neutral-pion calculation includes only the connected contributions. The disconnected contributions have not been evaluated. Dedicated lattice studies have examined sea-quark effects in polarizability calculations~\cite{Freeman:2014kka} and numerical treatment of disconnected quark loops~\cite{BARAL201964}. The results therefore do not represent the full neutral-pion polarizabilities in QCD.

\subsection{Pion mass extrapolation}

We now extrapolate the reconstructed polarizabilities to the physical pion mass. The lattice results are obtained at $m_\pi=370$, 600, 800, and 1100 MeV. We use weighted fits to determine the mass dependence and evaluate the fitted functions at $m_\pi=138$ MeV.

\begin{figure}
\centering
\includegraphics[width=0.45\textwidth]{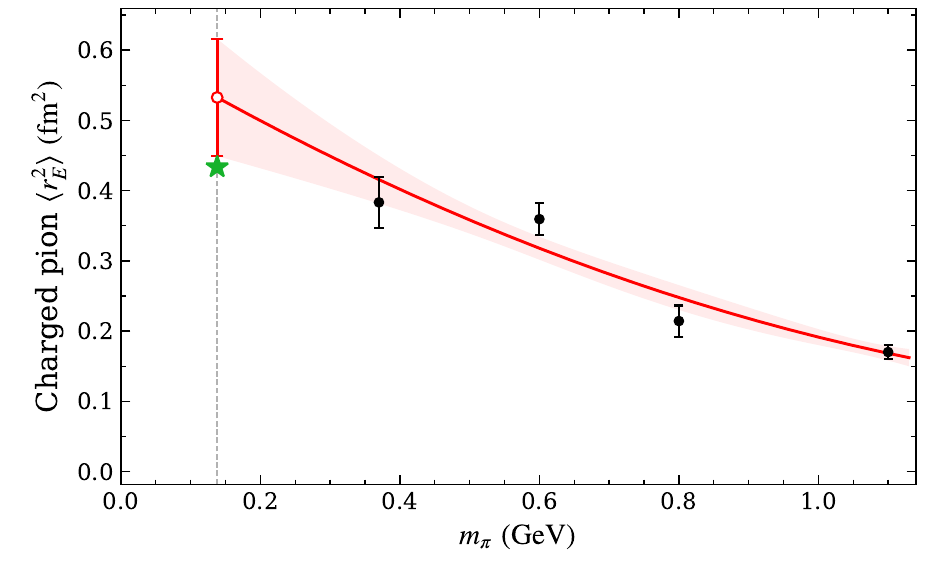}
\caption{
Charged-pion charge radius squared as a function of pion mass.  Filled points show the DDPM-based results from $F_\pi(Q^2)$ at the four simulated pion masses.  The solid curve is the weighted fit to a quadratic form in Eq.\eqref{eq:radiusmassfit},  and the shaded region indicates its propagated uncertainties. Open point denotes the extrapolated value at the physical point of $m_\pi=138$ MeV, while the green star indicates the PDG value.}
\label{fig:elastic_mass}
\end{figure}

We first consider the charged-pion elastic contribution. 
For the charged-pion charge radius, we use a quadratic form,
\begin{equation}
\langle r_E^2 \rangle
=a+b\,\mpi+c\,\mpi^2,
\label{eq:radiusmassfit}
\end{equation}
where $a$, $b$, and $c$ are fitting parameters.
The elastic contribution is simply related to the charge radius by $\alpha^{elas}_E=\alpha \langle r_E^2 \rangle /(3m_\pi)$, which leads to the form,
\begin{equation}
P_{elas}(\mpi)=\frac{a}{\mpi}+b+c\,\mpi.
\label{eq:elasticmassfit}
\end{equation}
Figure~\ref{fig:elastic_mass} shows the pion mass dependence of the charge radius squared. 
They are extracted from Eq.\eqref{eq:radius} using the DDPM-reconstructed form factor $F_\pi(Q^2)$ in Figure~\ref{fig:elastic_ff}. We observe some scatter in the extracted points, possibly associated with  the uncertainty in taking the derivative at the $Q^2=0$ point.  
We fit the data to Eq.\eqref{eq:radiusmassfit} and obtain its value at the physical pion mass $m_\pi=138$ MeV.
For reference, we also plot the experimental charge radius from PDG~\cite{PDG2024Review}, but caution that our result is subject to systematic uncertainties.

For the total electric polarizability of charged-pions in Eq.\eqref{eq:charged_e}, we add the two terms: elastic contribution $\alpha^{elas}_E=\alpha \langle r_E^2 \rangle /(3m_\pi)$, and the inelastic contribution.
For the total magnetic polarizability of charged-pions in Eq.\eqref{eq:charged_m}, we add the elastic contribution which is the negative of that in the electric case $\beta^{elas}_E= - \alpha^{elas}_E$, and the inelastic contribution. 
For the neutral-pion polarizabilities in Eq.\eqref{eq:neutral_e} and Eq.\eqref{eq:neutral_m}, there is no elastic contribution.
In all cases for the total, we use the form for the chiral extrapolation,
\begin{equation}
P_{total}(\mpi)=a+b\,\mpi+c\,\mpi^3.
\label{eq:massfit}
\end{equation}
The statistical errors of the DDPM-reconstructed results at $\bm q^2=0$
are used as weights in the fits.

\begin{figure}
\centering
\includegraphics[width=0.45\textwidth]{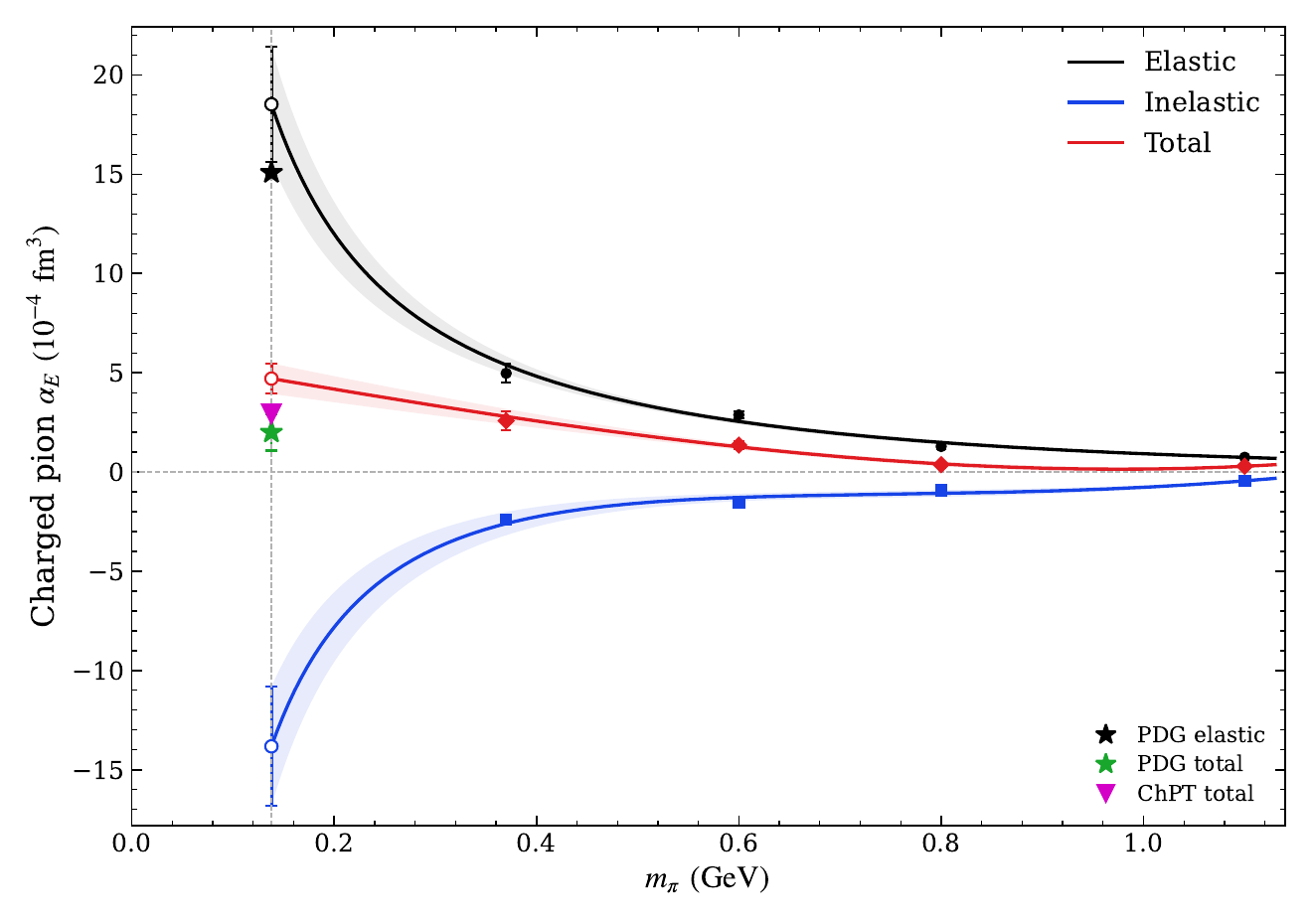}
\caption{Charged-pion electric polarizability versus pion mass from Eq.\eqref{eq:charged_e}. Black, blue, and red curves denote elastic, inelastic, and total contributions, respectively. Bands represent the propagated statistical uncertainty. 
Open points at 138 MeV denote extrapolations; stars and the triangle show the PDG and ChPT benchmarks identified in the legend.}
\label{fig:charged_e_mass}
\end{figure}

The charged-pion electric polarizability is displayed in Figure~\ref{fig:charged_e_mass}. 
The total polarizability is fitted using all four pion masses.
The inelastic mass dependence is then obtained by subtracting the fitted elastic contribution from the fitted total. We do not perform a separate mass fit for the electric inelastic contribution.
For reference, the extrapolated total is plotted together with the two-loop ChPT result~\cite{Gasser_2006}. Compared to the original result in Ref.~\cite{Lee:2023rmz}, the overall physics picture is unchanged: there is a large cancellation between the elastic and inelastic contributions, resulting in a mild pion mass dependence in the total.
The DDPM-based elastic value is slightly larger than the original estimate.
The DDPM analysis also uncovered an incompatibility issue in the $\bm q^2$ extrapolation at $m_\pi=800$ MeV, as detailed in Section~\ref{sec:800mev}. 

\begin{figure}
\centering
\includegraphics[width=0.45\textwidth]{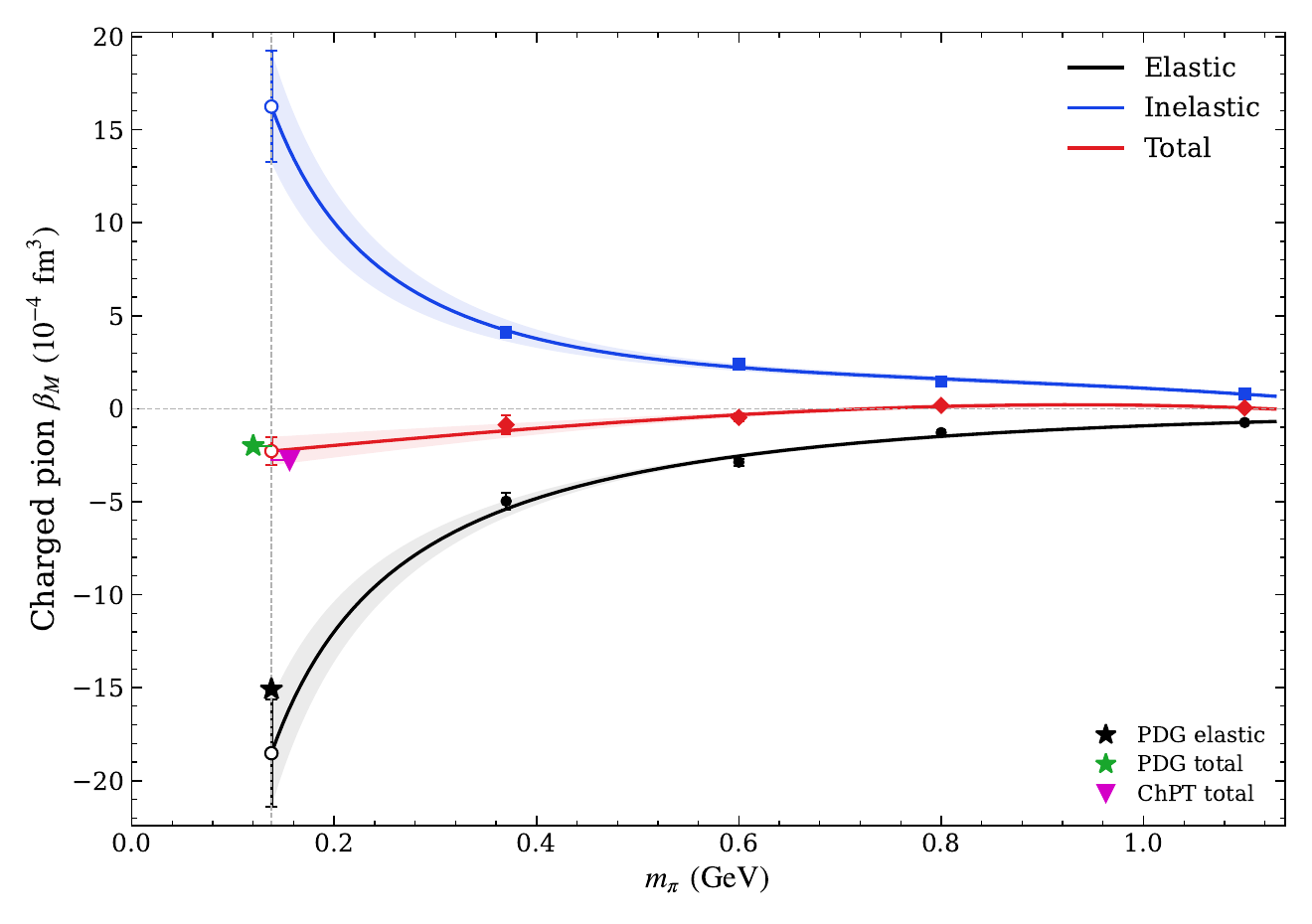}
\caption{Similar to Figure~\ref{fig:charged_e_mass}, but for charged-pion magnetic polarizability in Eq.\eqref{eq:charged_m}. 
The PDG total and ChPT total values are slightly offset horizontally for visibility.}
\label{fig:charged_m_mass}
\end{figure}

Figure~\ref{fig:charged_m_mass} shows the situation for the charged-pion magnetic polarizability. As in the electric case, the magnetic inelastic mass dependence is obtained by subtracting the elastic contribution from the fitted total. The overall picture remains largely the same as in the original result in Ref.~\cite{Lee2023ChargedMagnetic}.

For the neutral pion, we fit the electric and magnetic polarizabilities separately using Eq.~(\ref{eq:massfit}). All four simulated masses are included in the fits. Chiral extrapolations of the neutral-pion magnetic polarizability have also been investigated using background-field results~\cite{HeLeinweberThomasWang2020}. The sum of the two polarizabilities is obtained from the fitted electric and magnetic contributions.

\begin{figure}
\centering
\includegraphics[width=0.45\textwidth]{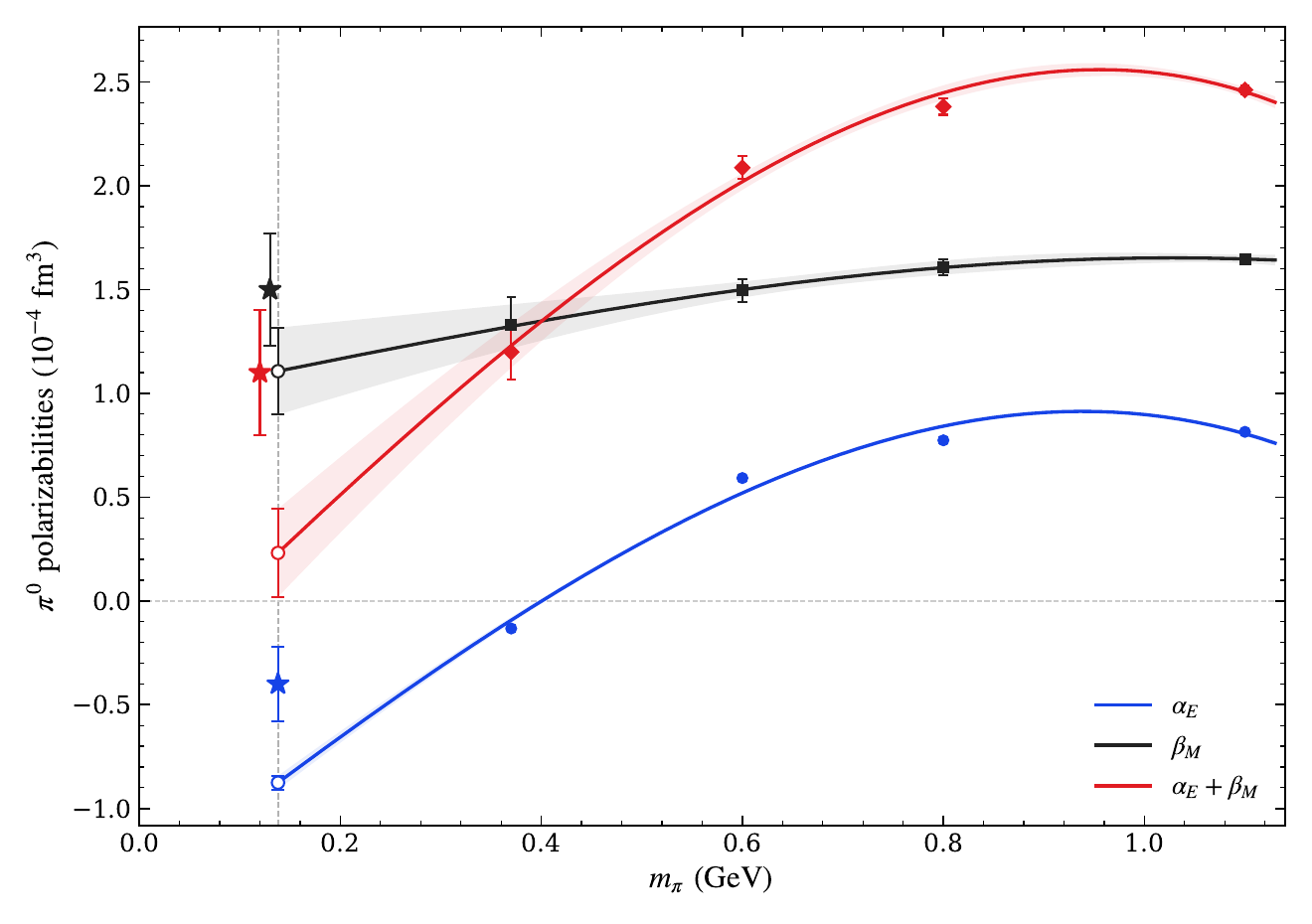}
\caption{Neutral-pion electric polarizability (blue) in Eq.\eqref{eq:neutral_e}, magnetic polarizability (black) in Eq.\eqref{eq:neutral_m}, and their sum (red) as functions of pion mass. The electric and magnetic curves are from separate four-mass weighted fits of the same functional form in Eq.\eqref{eq:massfit}; the sum is constructed from these two curves.
The ChPT values (stars) at  $m_\pi=138$ MeV are slightly offset horizontally for visibility.}
\label{fig:neutral_mass}
\end{figure}

Figure~\ref{fig:neutral_mass} shows the neutral-pion mass extrapolations. Two-loop ChPT studies of neutral-pion photon--photon amplitudes provide relevant theoretical benchmarks~\cite{GasserIvanovSainio2005}. The results at the physical pion mass are shown together with the reference values. Since the lattice input contains only connected diagrams, these extrapolations give the connected contributions and should not be interpreted as complete QCD results.

%


\begin{table*}
\centering
\caption{
Summary table for charge radius and electric and magnetic
polarizabilities of charged pions.
The values at the four pion masses are determined from the DDPM reconstruction in 
$\bm q^2$. They are extrapolated to the physical pion mass.
For reference, the extrapolated results are listed together with traditional fitting results and known
values.
All polarizabilities are in units of $10^{-4}\,{\rm fm}^3$.
}
\label{tab:charged}
\renewcommand{\arraystretch}{1.1}
\setlength{\tabcolsep}{3pt}

\begin{tabular}{c|cccc|c|c|c}
\hline
$m_\pi$ (MeV)
& $1100$ & $800$ & $600$ & $370$
& $138$
& Refs.~\cite{Lee:2023rmz,Lee2023ChargedMagnetic}
& Known value \\
\hline

$\langle r_E^2\rangle$ (fm$^2$)
& $0.170(10)$
& $0.214(22)$
& $0.359(23)$
& $0.383(36)$
& $0.533(83)$
& $0.40(5)$
& $0.434(5)$ (PDG) \\
\hline

$\alpha_E$ elastic
& $0.742(45)$
& $1.286(133)$
& $2.875(186)$
& $4.972(472)$
& $18.525(2.901)$
& $13.9(1.8)$
& $15.08(13)$ (PDG) \\

$\alpha_E$ inelastic
& $-0.453(27)$
& $-0.917(36)$
& $-1.519(25)$
& $-2.389(98)$
& $-13.819(3.001)$
& $-9.7(1.9)$
& \\

$\alpha_E$ total
& $0.290(55)$
& $0.369(138)$
& $1.356(190)$
& $2.583(486)$
& $4.706(767)$
& $4.2(5)$
& $\begin{array}{c}
2.93(5)\ \text{(ChPT)}\\
2.0(6)(7)\ \text{(PDG)}
\end{array}$ \\

\hline

$\beta_M$ elastic
& $-0.742(45)$
& $-1.286(133)$
& $-2.875(186)$
& $-4.972(472)$
& $-18.525(2.901)$
& $-13.9(1.8)$
& $-15.08(13)$ (PDG) \\

$\beta_M$ inelastic
& $0.787(5)$
& $1.462(12)$
& $2.408(26)$
& $4.112(66)$
& $16.245(3.000)$
& $12.4(1.9)$
& \\

$\beta_M$ total
& $0.045(46)$
& $0.177(135)$
& $-0.468(191)$
& $-0.860(485)$
& $-2.280(764)$
& $-1.4(5)$
& $\begin{array}{c}
-2.77(11)\ \text{(ChPT)}\\
-2.0(6)(7)\ \text{(PDG)}
\end{array}$ \\

\hline
\end{tabular}

\end{table*}


\begin{table*}
\centering
\caption{Similar to Table~\ref{tab:charged}, but for the neutral pion.}
\label{tab:neutral}

\renewcommand{\arraystretch}{1.1}
\setlength{\tabcolsep}{4pt}

\begin{tabular}{c|cccc|c|c|c}
\hline
$m_\pi$ (MeV)
& $1100$ & $800$ & $600$ & $370$
& $138$
&Ref.\cite{Lee2024Neutral}
& Known value \\
\hline

$\alpha_E^{\pi^0}$
& $0.815(11)$
& $0.774(13)$
& $0.592(12)$
& $-0.133(15)$
& $-0.876(33)$
& $-0.31(1)$
& $-0.40(18)$ (ChPT) \\

$\beta_M^{\pi^0}$
& $1.648(21)$
& $1.608(37)$
& $1.496(54)$
& $1.332(132)$
& $1.107(210)$
& $0.19(2)$
& $1.50(27)$ (ChPT) \\

$\alpha_E^{\pi^0}+\beta_M^{\pi^0}$
& $2.462(23)$
& $2.382(39)$
& $2.088(56)$
& $1.199(133)$
& $0.231(212)$
& $-0.12(2)$
& $1.10(30)$ (ChPT) \\
\hline
\end{tabular}

\end{table*}

Compared to the original result in Ref.~\cite{Lee2024Neutral}, the biggest change is in the magnetic polarizability $\beta_M$; it is now in better agreement with the ChPT value.
As a result, the sum $\alpha_E+\beta_M$ turns from negative to positive, in better agreement with the ChPT value, but still $2\sigma$ away. 
The neutral pion channel is more susceptible to disconnected contributions, see Figure~\ref{fig:diagrams}. 
 
The pion mass extrapolation remains subject to uncertainties associated with the choice of the fitting function. These uncertainties are not included in the statistical errors propagated from the lattice data. Although all four pion masses are used in the weighted fits, the available masses are considerably heavier than the physical pion mass. Additional lattice calculations at lighter pion masses would help constrain the mass dependence and improve the reliability of the extrapolation.

In Tables~\ref{tab:charged} and \ref{tab:neutral} we summarize the DDPM reconstruction numerical results at the four simulated pion masses and the extrapolations to the physical pion mass. The results are compared with the conventional fits reported in the original works and available reference values.

\section{Conclusion and Outlook}
\label{sec:sum}

In this work, we have applied generative machine learning in the 
analysis of pion electromagnetic polarizabilities from four-point functions in lattice QCD. 
A major source of systematic uncertainty in such calculations~\cite{Lee:2023rmz,Lee2023ChargedMagnetic,Lee2024Neutral} is the extrapolation of the $\bm q^2$ dependence of the four-point functions to the limit of $\bm q^2\to 0$ where the  polarizabilities are defined. Different fitting functions give different results, leading to large uncertainty spreads in the final results. To address this issue, we employ the so-called denoising diffusion probabilistic model (DDPM) in generative machine learning. 

DDPM provides an alternative to conventional fitting methods when only a few lattice momentum points are available. A few well-placed lattice points are enough for the reconstruction; sometimes as few as two points. Generally speaking, points at smaller momentum transfer carry more significance.
In our case, we have four nonzero momentum transfers denoted by $\bm q=(0,0,1)$, $(0,1,1)$, $(1,1,1)$, and $(0,0,2)$ at each of four  
pion masses $m_\pi=$370, 600, 800, and 1100 MeV. 
We subject the $\bm q^2$ dependence of our lattice input to DDPM-reconstruction, and then extrapolate the values obtained at $\bm q^2=0$ at each pion mass to the physical pion mass.

Instead of choosing a single fitting function, the method reconstructs the momentum dependence using information learned from several families of functions. 
We encountered three challenges in applying DDPM in this work.
The first is how to take derivatives of form factors to obtain charge radius from multiple families of training functions.
The second is to diagnose possible incompatibility issues in the input data by examining subsets.
The third is on how to deal with zero crossings. Examples and their resolutions are documented in the appendices.

The statistical errors in this work are propagated from the original lattice measurements. We use the median of the DDPM-generated curves for the central reconstruction and do not include the spread of these curves as a statistical error. However, the reconstruction still depends on the training functions and their parameter ranges. This dependence is not included in the quoted uncertainties. The method reduces the need to select a single fitting function, but it does not remove all assumptions about the momentum dependence. 

One possible area of improvement in the DDPM analysis is the chiral extrapolations. Right now, we use simple polynomial forms. One could perform a subsequent DDPM analysis on the pion mass dependence after the DDMP on the $\bm q^2$ dependence. There is a wealth of knowledge and functional forms to draw from ChPT as training functions for physics-informed extrapolations.

Future work should include lattice points at smaller momenta and lighter pion masses. Dynamical lattice simulations and the inclusion of disconnected contributions would provide more complete physical results. 
To this end, we plan to apply DDPM on more datasets based on dynamical two-flavor nHYP-smeared ensembles at $m_\pi=315$ MeV and $m_\pi=220$ MeV and several volumes to reach smaller momenta~\cite{Luke:2026wkn}. The disconnected diagrams are more challenging. 
Future evaluations could draw on established diluted all-to-all propagator estimators~\cite{Foley2005AllToAll}, subtraction methods~\cite{BARAL201964}, distillation-based quark-field methods~\cite{Peardon2009Distillation}, and stochastic Laplacian-Heaviside (LapH) smearing~\cite{Morningstar2011StochLapH}.
We are also exploring a method based on  LapH-smeared currents~\cite{lee2026vector}.
 
With improved lattice data and further tests of the reconstruction, the DDPM may provide a useful numerical method for studying hadronic observables in lattice QCD.

\begin{acknowledgments}
This work was supported in part by U.S. Department of Energy under  Grant~No.~DE-FG02-95ER40907. Most of the calculations were carried out on Perlmutter GPUs at DOE-sponsored NERSC. 
\end{acknowledgments}

\appendix


\section{Machine-learning implementation}
\label{app:DDPM}
Some details of the DDPM method follow the work on gravitational form factors~\cite{Alharazin_2026}, while the implementation is adapted to the pion polarizability calculations in this work. The training data are generated
from the four function families described in
Sec.~\ref{sec:fun}, with 50,000 curves from each family,
giving a total of 200,000 curves. Each curve is evaluated
on a uniform grid of 200 points over the squared-momentum
range from 0 to 2 GeV$^2$. The four families are assigned
equal weights, since there is not significant evidence to
make any function with larger weight at this stage. Different
weights may be considered after further tests of the
training functions. We use AdamW with a learning rate of
$10^{-4}$, weight decay of $10^{-4}$, and batch size of 256.
The training consists of 100 epochs, including a two-epoch
warmup followed by cosine learning-rate decay. Gradient
clipping is set to 1, the exponential-moving-average factor
is 0.9999, and the random seed is fixed at 7.

The lattice data are used to constrain the reconstructed
curves at the available momentum points. For the elastic
form factor, the condition $F_\pi(0)=1$ is imposed, and
the derivative at $Q^2=0$ is extracted from the reconstructed
curve. For the inelastic contributions, the reconstructed
values at $\qq=0$ are used directly. We take the pointwise
median of $N=1000$ generated curves and use $M=100$
sets of lattice inputs to propagate the statistical errors.
Tests with larger values of $M$ and $N$ show very little
change in the extracted results and their uncertainties.
Further increases are therefore unnecessary for the
present calculation.

We have tested different combinations of training function
families. The spread of the generated curves is smallest
when only one family is used and generally increases when
more families are included. However, a smaller spread does
not necessarily mean a more accurate reconstruction.
Including different functions allows a wider range of
momentum dependence, which is one of the reasons for using
the DDPM rather than a single fitting function. The spread
of the generated curves therefore cannot be taken as the
statistical error of the physical result. Instead, we use
the statistical errors propagated from the original lattice
data, as described in the main text. The uncertainty
associated with the choice of training functions is not
included in the quoted errors. This limitation is
particularly relevant near zero momentum, where no lattice
data are available.

\section{Derivative-window analysis}
\label{app:deriv}

The elastic contribution to the charged-pion polarizability requires the derivative of the electromagnetic form factor at zero momentum. We determine this derivative by fitting the DDPM median form factor with a quadratic function,
\begin{equation}
F_\pi(Q^2)=A+BQ^2+C(Q^2)^2,
\label{eq:elastic_quadratic}
\end{equation}
where the coefficient $B$ gives the derivative at $Q^2=0$. Since the reconstructed form factor is available on a dense momentum grid, we examine the dependence of the derivative on the fitting range by varying its upper limit $Q^2_{\max}$.

\begin{figure*}
\centering
\includegraphics[width=0.93\textwidth]{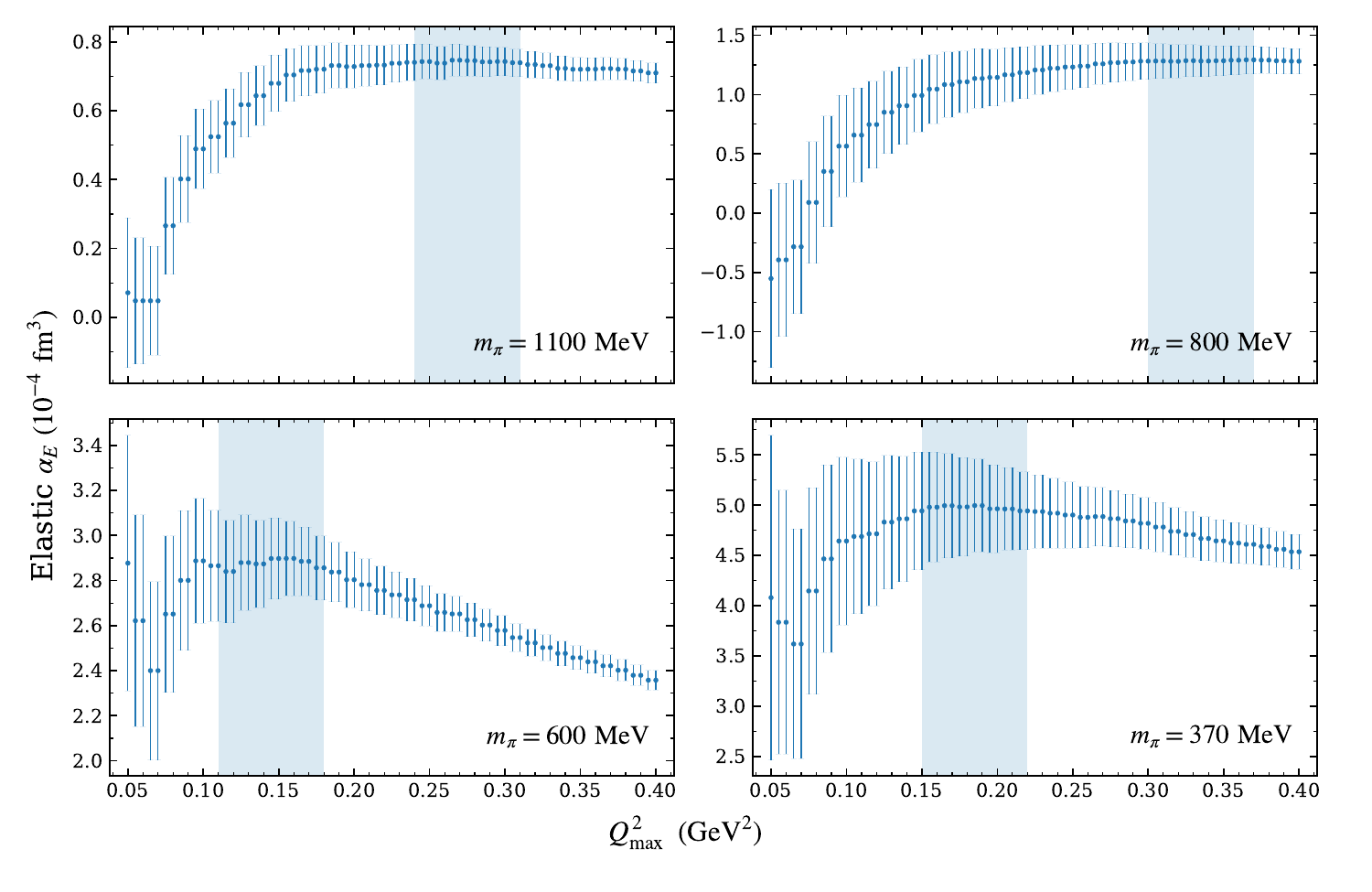}
\caption{
Dependence of the charged-pion elastic electric polarizability on the upper endpoint $Q^2_{\max}$ of the local form-factor fit, for $m_\pi=1100$ and 800 MeV (upper panels) and 600 and 370 MeV (lower panels). Points and error bars denote the extracted values and propagated lattice uncertainties. Shaded vertical bands mark the selected fitting-window ranges.
}
\label{fig:derivative}
\end{figure*}

Figure~\ref{fig:derivative} shows the fitted derivatives as functions of $Q^2_{\max}$ for the four pion masses. At small $Q^2_{\max}$, the fitted derivatives fluctuate considerably because the quadratic fit is sensitive to small variations in the reconstructed curve. As $Q^2_{\max}$ increases, the fluctuations become smaller and a relatively stable region appears. At still larger $Q^2_{\max}$, the derivative changes more smoothly, but the fit becomes increasingly influenced by the form factor at higher momentum. A smoother derivative therefore does not necessarily provide a better estimate at $Q^2=0$.

We select the intermediate regions where the initial fluctuations have become small without extending the fit too far into the higher-momentum region. For consistency, the same window width of $0.07\,{\rm GeV}^2$ in $Q^2_{\max}$ is used for all four pion masses. The selected ranges of $Q^2_{\max}$ are $[0.15,0.22]\,{\rm GeV}^2$ for 370 MeV, $[0.11,0.18]\,{\rm GeV}^2$ for 600 MeV, $[0.30,0.37]\,{\rm GeV}^2$ for 800 MeV, and $[0.24,0.31]\,{\rm GeV}^2$ for 1100 MeV. These windows are indicated by the shaded regions in Fig.~\ref{fig:derivative}. Their positions differ because the reconstructed form factors have different momentum dependences.

For each pion mass, we take the mean of the fitted derivatives over the selected range of $Q^2_{\max}$ as the final derivative at $Q^2=0$. This derivative is used to determine the charge radius through Eq.~(\ref{eq:radius}). The same fitting and averaging procedure is repeated for the resampled lattice inputs to propagate their statistical errors to the elastic polarizability. The window analysis provides a check on the stability of the derivative extraction, although it does not remove the possible dependence on the reconstructed form-factor shape.

\section{Diagnostic of lattice input at 800 MeV}
\label{app:800diagnostic}

The charged electric inelastic contribution at
$m_\pi=800$ MeV shows a different momentum dependence
from those at the other pion masses, as observed in
Figure~\ref{fig:charged_einel}. To examine the sensitivity of the DDPM
reconstruction to the lattice input, we repeat the
calculation using different subsets of the four
available lattice momenta.

The four momentum points are labeled as
\begin{equation}
\begin{aligned}
1 &: \boldsymbol{q}=(0,0,1),\\
2 &: \boldsymbol{q}=(0,1,1),\\
3 &: \boldsymbol{q}=(1,1,1),\\
4 &: \boldsymbol{q}=(0,0,2).
\end{aligned}
\end{equation}
The corresponding squared momenta are
$\boldsymbol{q}^{\,2}
=0.3663,\ 0.7327,\ 1.0990,$ and
$1.4654\ \mathrm{GeV}^2$, respectively.

The elastic contribution is reconstructed from all
four lattice form-factor points using $Q^2$ in every
calculation. Only the lattice points used for the
inelastic contribution, expressed in terms of
$\boldsymbol{q}^{\,2}$, are varied. The same trained
DDPM and reconstruction procedure are used throughout.
For each subset, the central reconstruction is obtained
from the pointwise median of $N=1000$ generated curves.
The uncertainties are propagated through $M=100$
lattice resamplings. The spread of the generated
DDPM curves is not included as an additional
uncertainty.

To examine the dependence on the lattice input,
we consider all four possible three-point subsets
and all six possible two-point subsets.

%
To save space, we only display the reconstructed polarizabilities
at $\bm q^2=0$ in Table~\ref{tab:800subsets}, not the full reconstructed $\bm q^2$-curves. 
The elastic contribution is unchanged because
the same four-point form-factor reconstruction
using $Q^2$ is used for every subset. The total
polarizability is obtained by adding the elastic
and inelastic contributions, with its uncertainty
determined from the lattice resamplings. 

\begin{table}
\centering
\caption{Charged electric polarizability at
$m_\pi=800$ MeV for different momentum subsets,
in units of $10^{-4}\,\mathrm{fm}^3$.}
\label{tab:800subsets}

\begin{tabular}{cccc}
\hline
Subset & Elastic & Inelastic & Total \\
\hline
(1234) & 1.286(133) & -1.584(81) & -0.298(156) \\
(123)  & 1.286(133) & -1.398(75) & -0.112(153) \\
(124)  & 1.286(133) & -1.543(68) & -0.257(149) \\
(134)  & 1.286(133) & -2.574(79) & -1.288(155) \\
(234)  & 1.286(133) & -1.801(92) & -0.515(162) \\
(12)   & 1.286(133) & -1.537(69) & -0.251(150) \\
(13)   & 1.286(133) & -2.113(35) & -0.827(138) \\
(14)   & 1.286(133) & -1.808(29) & -0.522(136) \\
(23)   & 1.286(133) & -1.519(59) & -0.233(145) \\
(24)   & 1.286(133) & -0.917(36) & 0.369(138) \\
(34)   & 1.286(133) & -0.132(8) & 1.154(133) \\
\hline
\end{tabular}

\end{table}

We see that out of all the possibilities, only the (2,4) subset 
gives positive value compatible with the values at the other three masses.
The (3,4) subset also gives a positive value, but is ruled out because it does not constrain the behavior at small $\bm q^2$.
All other subsets give negative values for the total.
Based on the sensitivity tests, we think there is a possible issue with the $\bm q=(1,1,1)$ result in the lattice input. 
The (2,4) result is used in the mass extrapolation presented in Figure~\ref{fig:charged_e_mass}. 

These comparisons show that the low-momentum
continuation at $m_\pi=800$ MeV is sensitive
to the choice of lattice momentum points.
The quoted uncertainties describe the propagation
of lattice statistical errors for a fixed subset
and do not include the variation among different
subsets. The subset comparison therefore provides
a diagnostic of the input dependence, but does
not by itself establish which subset gives the
unique physical continuation. 
We include it based on the assumption that polarizability is a smooth function of pion mass.

\begin{figure*}
\centering
\includegraphics[width=\textwidth]{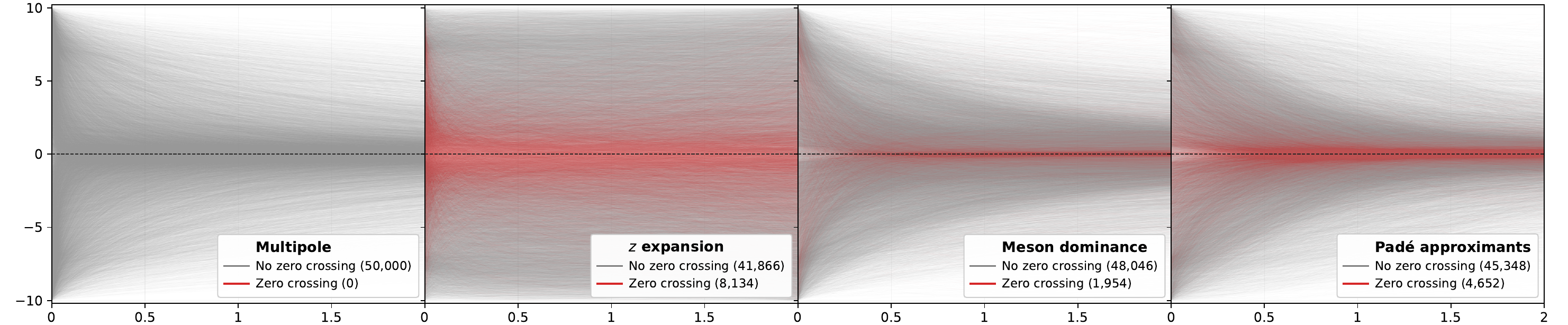}
\caption{
Synthetic training curves from the four function families.
Gray curves do not cross zero, while red curves cross zero
within $0\leq\mathbf{q}^2\leq2~\mathrm{GeV}^2$.
Each family contains 50,000 curves. The panels show,
from left to right, the multipole, $z$-expansion,
meson-dominance, and Pad\'e-approximant families.
}
\label{fig:training_zero_crossing}
\end{figure*}

\section{Zero crossings in neutral-pion magnetic reconstruction}
\label{app:neutral_shift}

The neutral-pion magnetic lattice data show an increasing number of negative values as the pion mass decreases. At $m_\pi=1100$ MeV, all four lattice points are positive. At 800 MeV, only one point has a small negative value, while two points are negative at 600 MeV. At 370 MeV, three of the four points are negative, with two reaching approximately $-0.161$ and $-0.142$. Such a large change of sign is unusual among the lattice data considered in this work, particularly since most of the DDPM training curves do not cross zero. The zero-crossing properties of the four training function families are shown in Fig.~\ref{fig:training_zero_crossing}.

The four training function families have different
zero-crossing properties, as shown in
Fig.~\ref{fig:training_zero_crossing}. Among the 200,000 training curves, the multipole family
has no zero crossings, while the meson-dominance,
$z$-expansion, and Pad\'e families contain 1,954 (3.91\%),
8,134 (16.27\%), and 4,652 (9.30\%) zero-crossing curves,
respectively. In total, 14,740 curves (7.37\%) cross zero.
The training set also contains positive and negative curves
with approximately symmetric behavior about zero. For
negative lattice inputs, different curves may give similar
values at the measured points but behave differently toward
$\bm q^2=0$. Some become more negative, while others
cross zero and approach positive values. With only four
lattice points, this ambiguity can affect the reconstructed
magnetic polarizability.

To examine this effect, we add a constant shift to the
magnetic lattice inputs before reconstruction and subtract
the same shift from the reconstructed median curves afterward.
Only the magnetic channel is shifted. At $m_\pi=370$ MeV,
the minimum lattice value is approximately $-0.1610$.
A shift of $0.165$ makes all four input values positive.
The reconstructed magnetic polarizability changes from
$\beta_M=1.4101(1321)$ without the shift to
$\beta_M=1.3322(1316)$ with the shift, a difference of
$0.078$ between the central values.

At $m_\pi=600$ MeV, the minimum lattice value is
approximately $-0.0493$. A shift of $0.05$ changes
$\beta_M$ from $1.3398(526)$ to $1.4957(542)$,
a difference of $0.156$. Thus, the smaller negative
input at 600 MeV produces a larger change than at
370 MeV. No shift is applied at 800 or 1100 MeV.
All magnetic polarizabilities and shifts are given
in units of $10^{-4}\,{\rm fm}^3$.

The training curves are approximately symmetric about zero, but only a small fraction of them cross zero. Thus, the
positive and negative curves are well represented, while
curves that change sign are much less common. This may
affect the reconstruction when the magnetic lattice data
contain negative values. For example, a curve passing
through negative lattice points may rise toward a positive
value at zero momentum. Such behavior requires a zero
crossing, whereas the much larger number of curves that
preserve their sign may influence the reconstructed
momentum dependence. As a result, the DDPM median may
be affected by the unequal representation of these
different behaviors in the training data.

By shifting the lattice inputs above zero, the same
low-momentum behavior can be represented without requiring
the training curves to cross zero. The shift is then
subtracted from the reconstructed curves to recover the original values. This provides a possible explanation
for the differences between the shifted and unshifted
results. The shift does not change the lattice information,
but changes how the input data are represented relative
to the training distribution.

\bibliography{xrefs}
\end{document}